\documentstyle[11pt,newpasp,twoside,epsf]{article}
\def\edcomment#1{\iffalse\marginpar{\raggedright\sl#1\/}\else\relax\fi}
\reversemarginpar

\begin{document}
\title{Stellar Flares and Coronal Structure}
\author{Manuel G\"udel}
\affil{Paul Scherrer Institut, 5232 Villigen PSI, Switzerland}

\begin{abstract}
Coronal structure and coronal heating are intimately related in magnetically
active stars. Coronal structure is commonly inferred from radio
interferometry and from eclipse and rotational modulation studies. We discuss to
what extent  flares may be responsible for coronal structure
and global observable properties in magnetically active stars.
\end{abstract}

\section{Introduction}

What are stellar coronae? The answer to this simple-minded question is deeply rooted in 
the interplay between magnetic field generation and energy release in stellar atmospheres. The nature 
of this interplay is surprisingly difficult to grasp.
While observables such as coronal plasma temperatures, emission measures, electron densities,
non-thermal particle densities and in cases magnetic field strengths  have painstakingly been
measured, monitored, and correlated, our question remains one of understanding cause and
effect in a complicated array of  interactions. There is  little doubt that
magnetic fields not only confine coronal plasma  but also heat this plasma
once they interact due to shearing and twisting as a consequence of surface convective
motions. Coronal heating thus intimately depends on magnetic structure; conversely, magnetic 
reconnection determines the geometry of radiating magnetic structures.

\section{Coronal Structure: Magnetospheres, Flares, Active Regions}

\subsection{Radio Interferometric Imaging of Stellar Coronae}

Radio Very Long Baseline Interferometry (VLBI)  has produced some of the best evidence of 
stellar coronal structure related to non-thermal electrons around active stars, on 
milliarcsecond (mas) scales. Since the electrons
are trapped in magnetic fields, they outline their overall structure. The achievable 
resolution is still only of order of a  stellar diameter for nearby active stars or 
of the intrabinary distance for the closest RS CVn and Algol binary systems. 

RS CVn and Algol-type binaries often reveal a  bi-modal compact core + extended halo radio structure,  
the halo being of a total size comparable to the binary system size (e.g., Mutel et al. 1985). 
The extended structure is resolved  in two oppositely polarized radio lobes in Algol that are aligned with
the rotation axis (Mutel et al. 1998; Fig. 1b). The large-scale structure seems to 
be based on globally ordered magnetic fields as suggested by  polarization gradients  
(Beasley \& G\"udel 2000). Other stellar classes exhibiting large-scale magnetospheres 
include pre-main sequence and magnetic chemically peculiar Bp/Ap stars 
(Phillips \& Lestrade 1988; Phillips, Lonsdale, \& Feigelson 1991, 1993; 
Andr\'e et al. 1991, 1992; Smith et al. 2003) with sizes up to 10$R_*$.  

\begin{figure}[t!]
\hbox{\plotfiddle{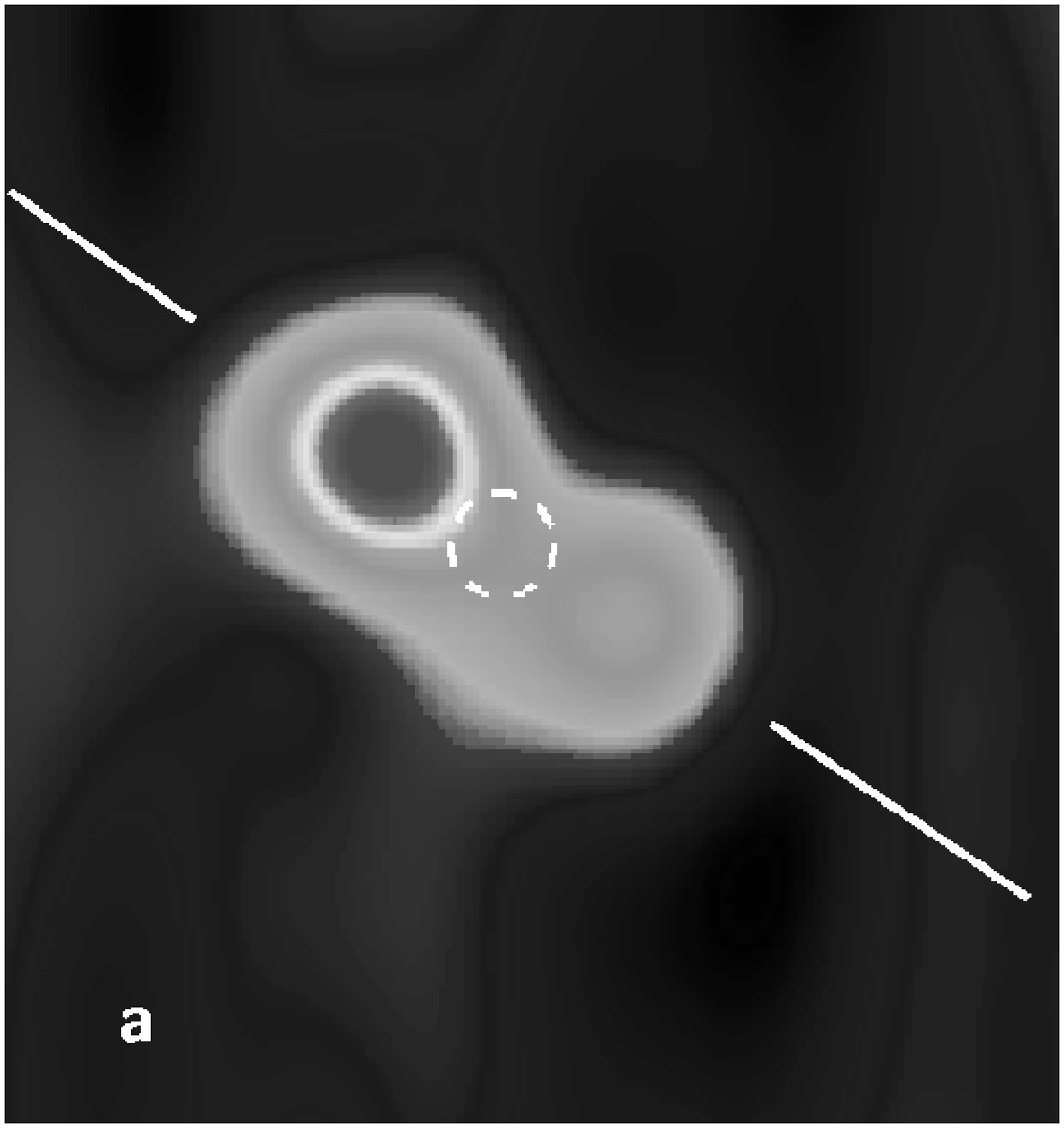}{0cm}{0}{32.5}{32.5}{-390}{-232}\plotfiddle{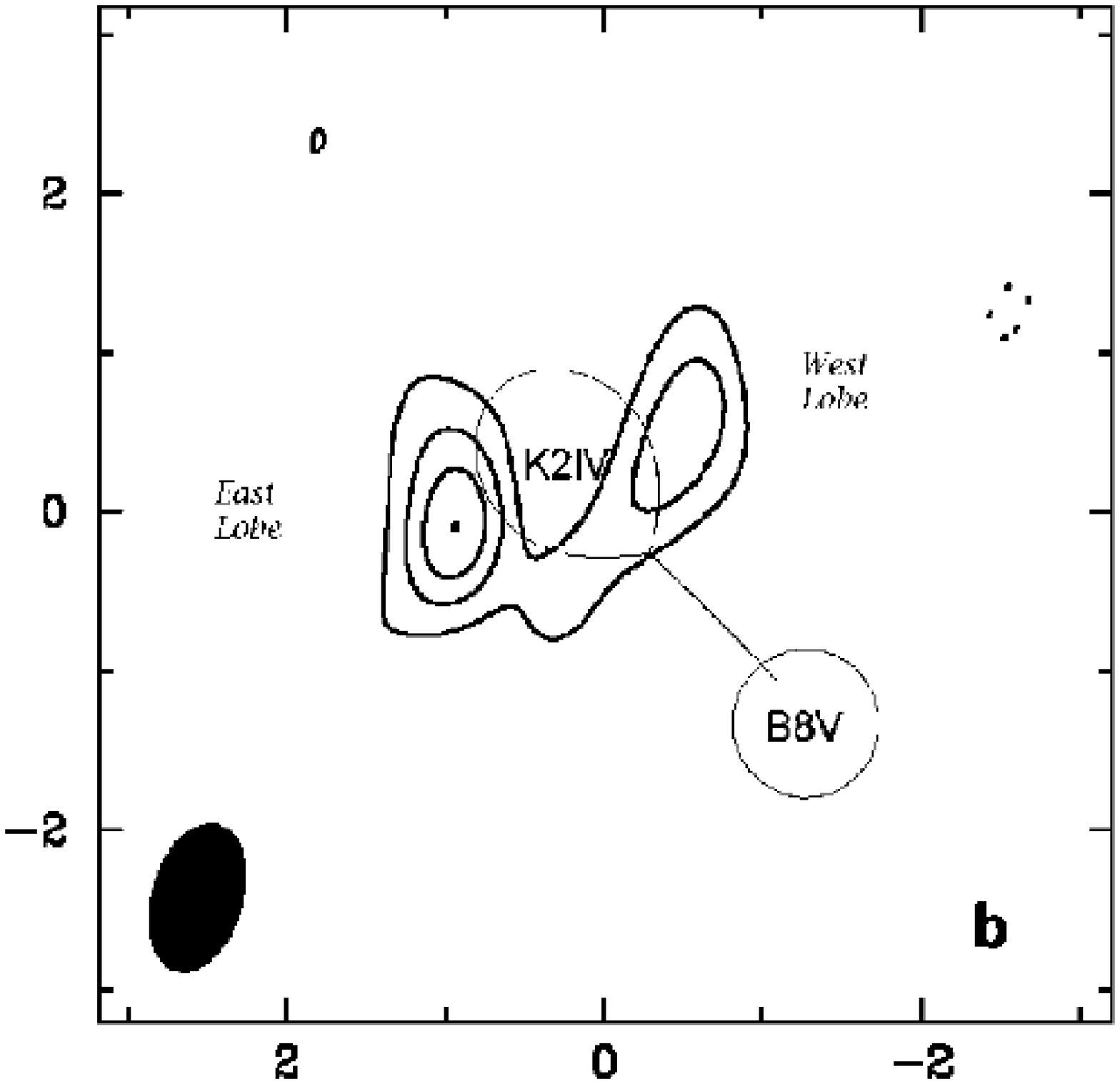}{0cm}{0}{37}{37}{-594}{-242}}
\hbox{\plotfiddle{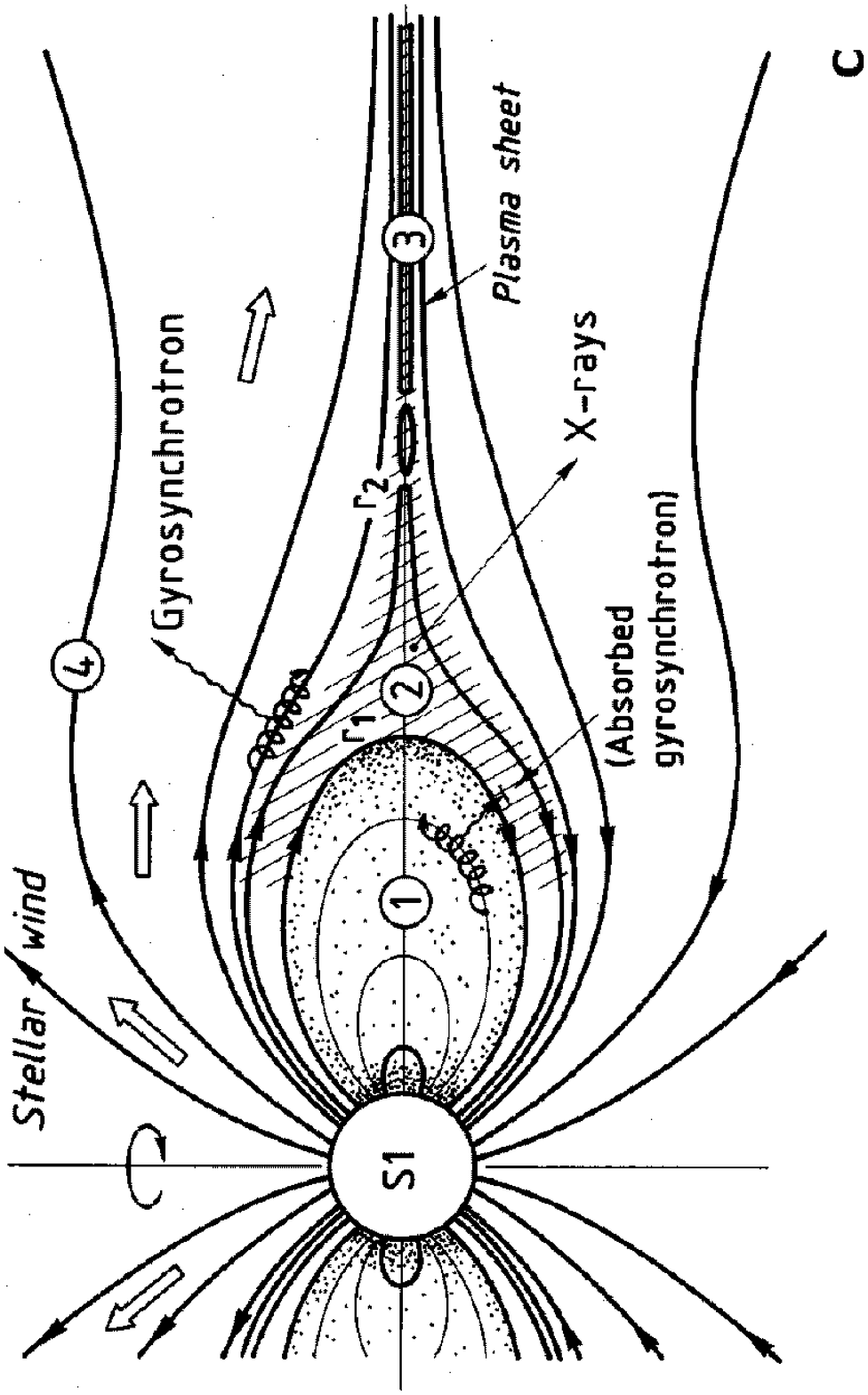}{0cm}{-90}{23}{23}{-390}{-190}\plotfiddle{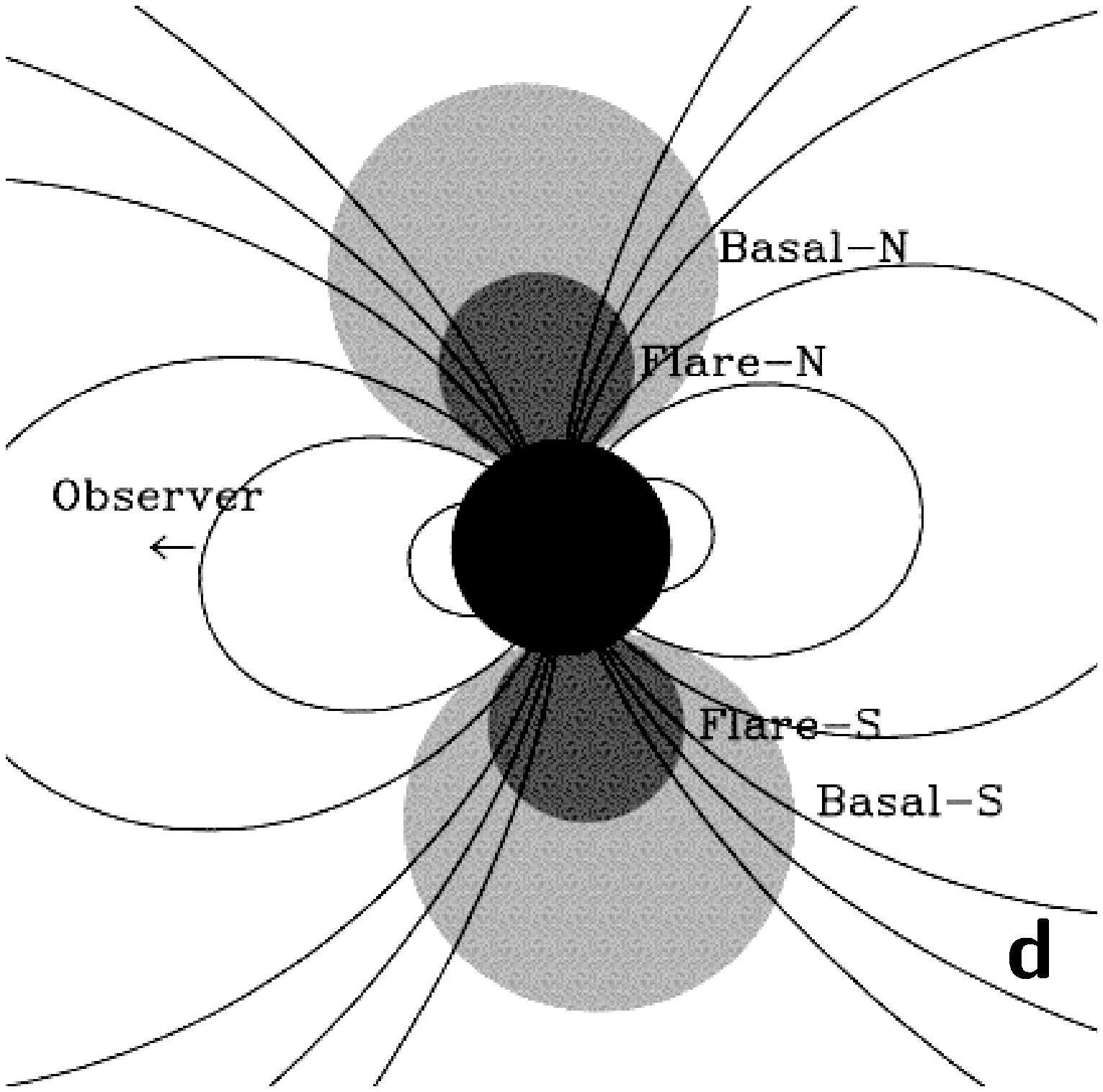}{0cm}{0}{33}{33}{-580}{-420}}
\vskip 11.5truecm
\caption{Radio coronal  structure: Top left: VLBI image of UV Cet (resolution 
$\approx 0.8$~mas; after Benz et al. 1998). Top right: VLBI image of Algol
(Mutel et al. 1998; units: mas). Bottom left: Model for a radio magnetosphere of a
young star (Andr\'e et al. 1988). Bottom right: Magnetospheric model
after Mutel et al. (1998).}
\vskip -1truecm
\end{figure}

The spatial and polarimetric boundary conditions are satisfied by 
global dipole-like magnetospheres  somewhat resembling the Earth's Van Allen 
belts (e.g., Andr\'e et al. 1988; Morris, Mutel, \& Su 1990; 
Linsky, Drake, \& Bastian 1992; Figure 1c).  Stellar winds escaping at higher latitudes could
draw 
the magnetic field lines into a current sheet configuration in the equatorial plane.  
Particles accelerated in that region travel back and are trapped in the 
equatorial magnetospheric cavity. Additional, dynamic ingredients may be required:  
The compact cores sometimes seen during enhanced flux episodes are likely to be 
expanding flares that evolve into extended halos where the  low-density population of 
accelerated electrons slowly cool on time scales of days (Mutel et al. 1985).

VLBI techniques have been more demanding for  dwarf stars, but 
spatially resolved observations of some M dwarfs also reveal 
coronae of a size much larger than the stellar diameter 
(Alef, Benz, \& G\"udel 1997; Pestalozzi et al. 2000). For example, the dMe star UV Cet B is surrounded 
by a pair of giant synchrotron lobes, with sizes up to 
$2.4\times 10^{10}$~cm and a separation of 4$-$5 stellar radii along the 
putative rotation axis of the star, suggesting very extended magnetic structures above the 
magnetic poles (Benz, Conway, \& G\"udel 1998; Fig. 1a). In this case, trapped high-energy electrons  
radiate most efficiently in the converging magnetic fields over the polar
regions (see also Mutel et al. 1998 for a similar model for the  magnetic field of Algol, Fig. 1d). 

The prime contribution of VLBI to  coronal physics has thus been in the recognition 
of extended, globally structured magnetic fields containing a variable number of
high-energy electrons. The origin of the latter is not definitively known, and 
our knowledge of  the connectivity of the magnetospheres to the surface magnetic 
fields is model-dependent.

\subsection{Eclipses}

The most productive methods of coronal structure mapping are based on 
(usually non-unique) image reconstructions from eclipse light curves 
(White et al. 1990 = W90; Siarkowsi 1992;  Schmitt 1996; G\"udel et al. 2003a).

Studies of extremely active binaries support a  bi-modal thermal and geometric structure,  
somewhat reminiscent of structure inferred from VLBI and
clearly unlike the non-flaring solar corona (Walter et al. 1983 = W83; 
White et al. 1986 = W86; Stern et al. 1992; Ottmann 1993, 1994; Ottmann, Schmitt, \& 
K\"urster 1993 = O93):  i) Judged from rather shallow
or absent X-ray eclipses in some RS CVn or Algol-type binaries, 
a significant fraction of the magnetically enclosed X-ray emitting 
plasma must be extended on scale heights $\ga R_*$; these structures are often 
associated with subgiant components (e.g., in AR Lac or Algol). 
They seem to be predominantly very hot in X-rays ($\gg 10$~MK), reminiscent of flaring plasma. 
The non-detection of radio eclipses (Doiron \& Mutel 1984 for AR Lac; van den Oord et al. 
1989 and Mutel et al. 1998 for Algol; Alef et al. 1997 for YY Gem) similarly suggests an extended 
magnetosphere also detected and resolved by VLBI (\S 2.1). 
ii) A second type of X-ray structure found in both evolved  and main-sequence binaries
shows high pressure (up to several 100~dyne~cm$^{-2}$) and small, ``solar-like'' 
scale  heights. Although they may be cooler than the extended structure, their pressure 
is reminiscent of flares.

\begin{figure}[t!]
\vskip -0.0truecm
\hbox{\plotfiddle{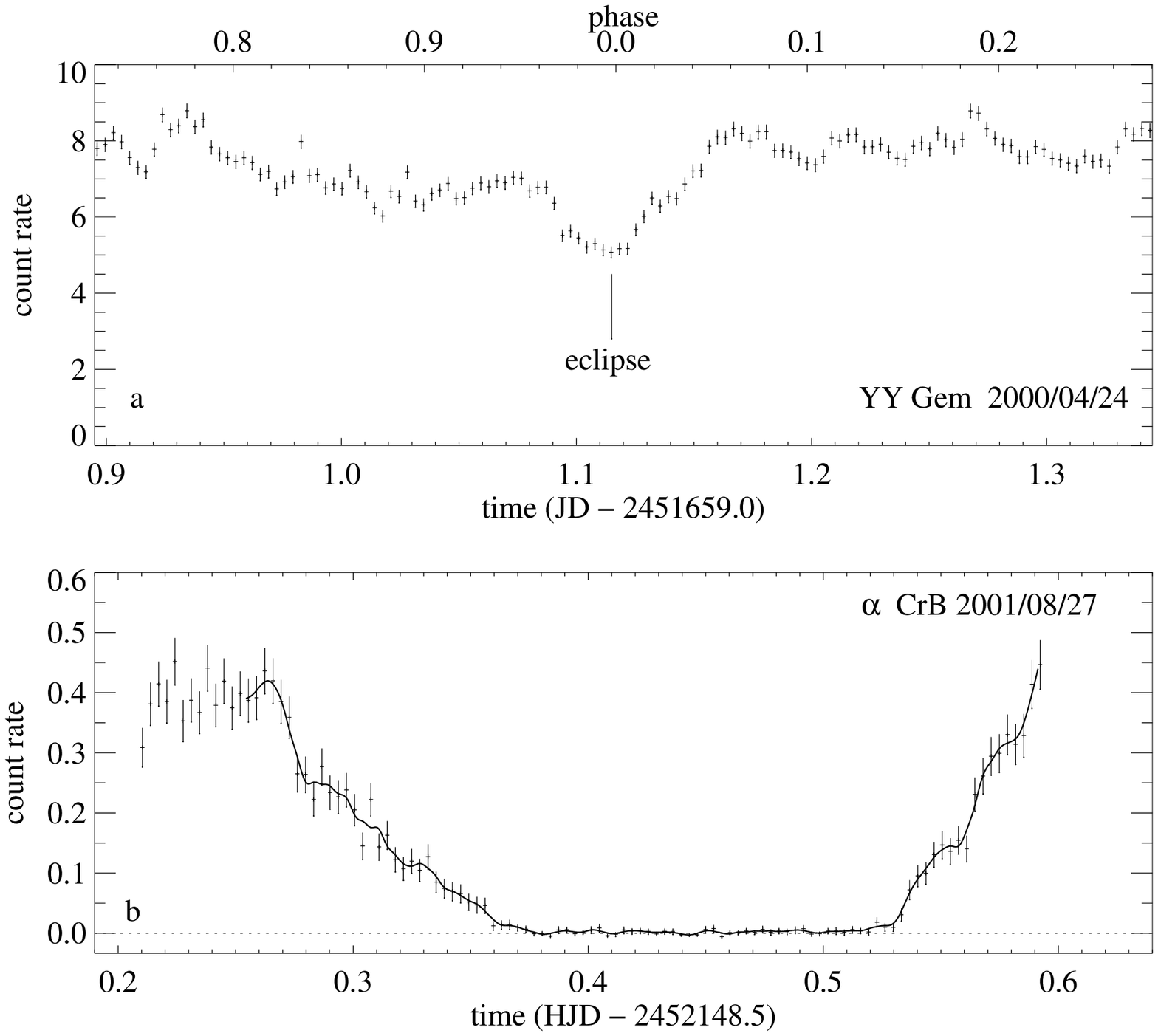}{0cm}{0}{47}{47}{-330}{-230} }
\hbox{\plotfiddle{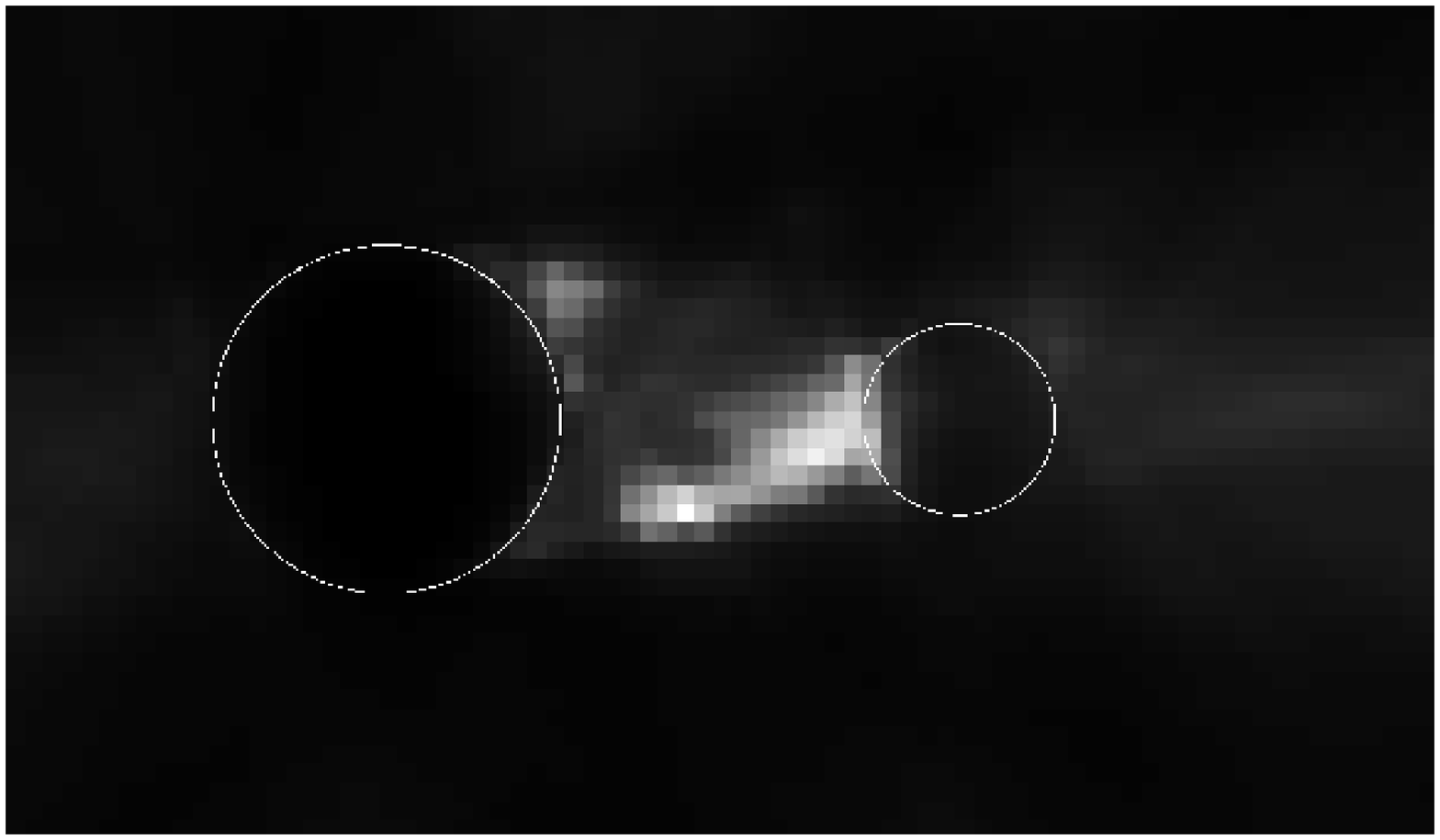}{0cm}{0}{40}{40}{-388}{-446.5} \plotfiddle{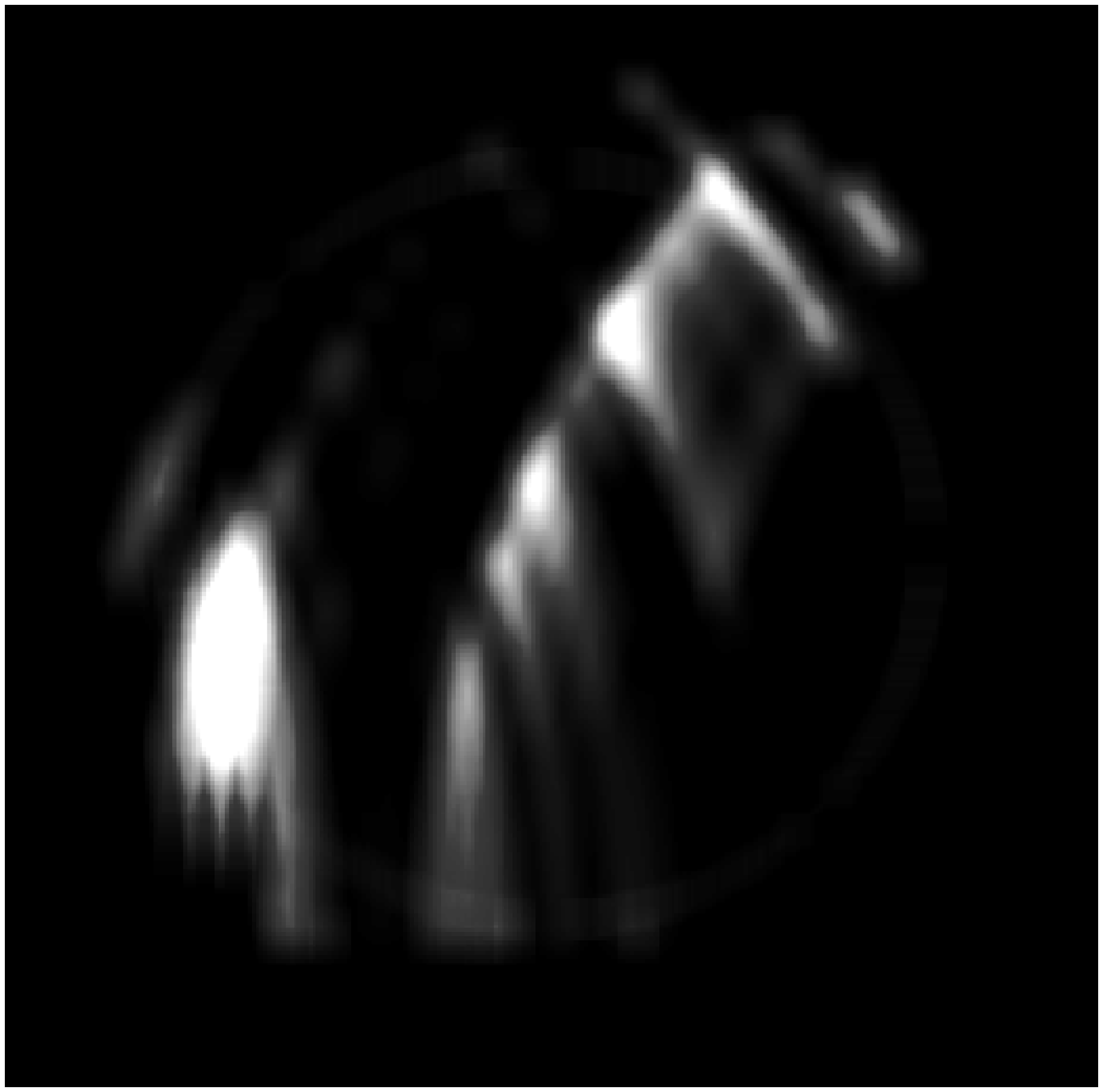}{0cm}{0}{26}{26}{-535.4}{-390.4}}
\hbox{\plotfiddle{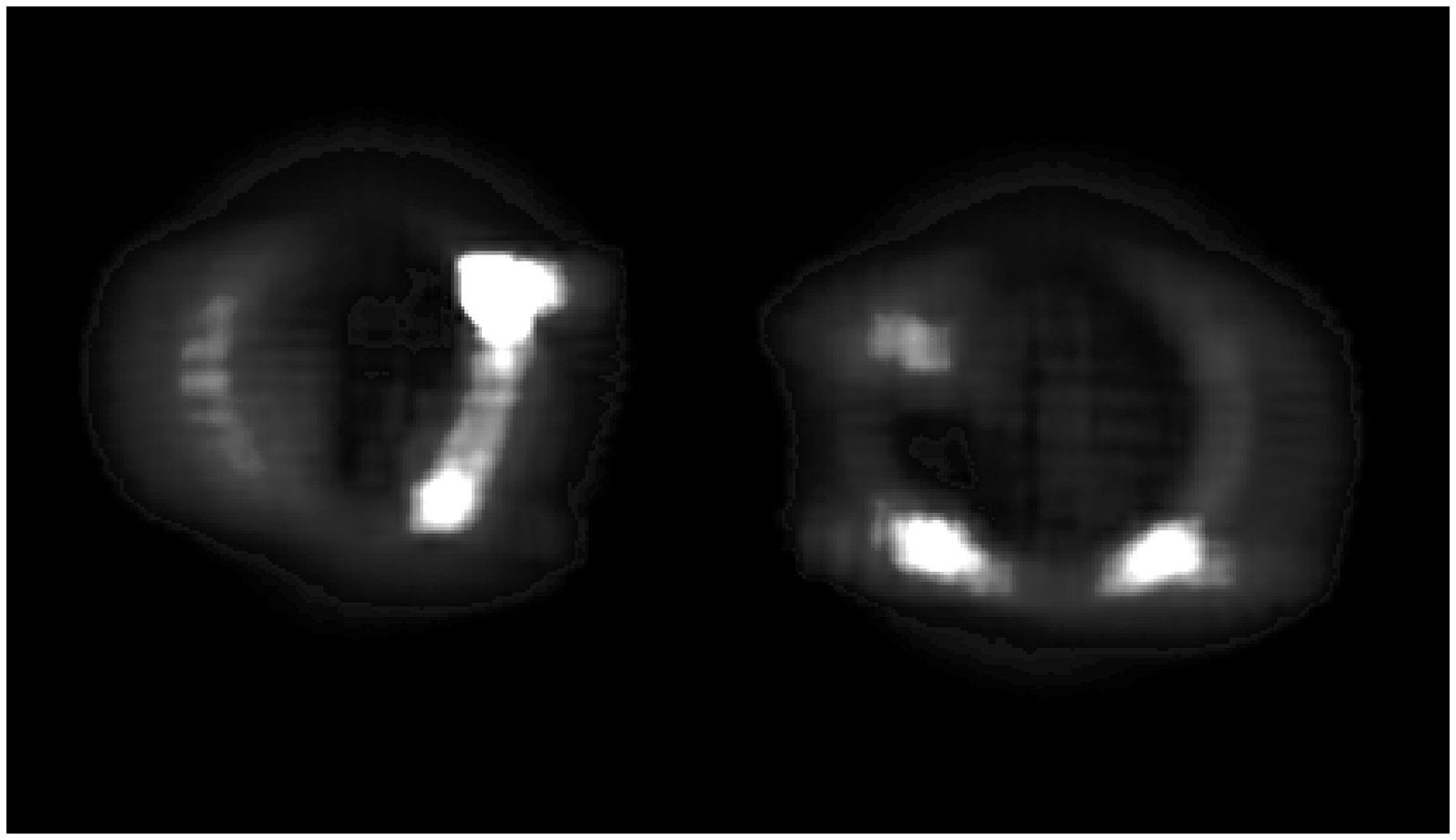}{0cm}{0}{44.5}{44.5}{-398}{-607} \plotfiddle{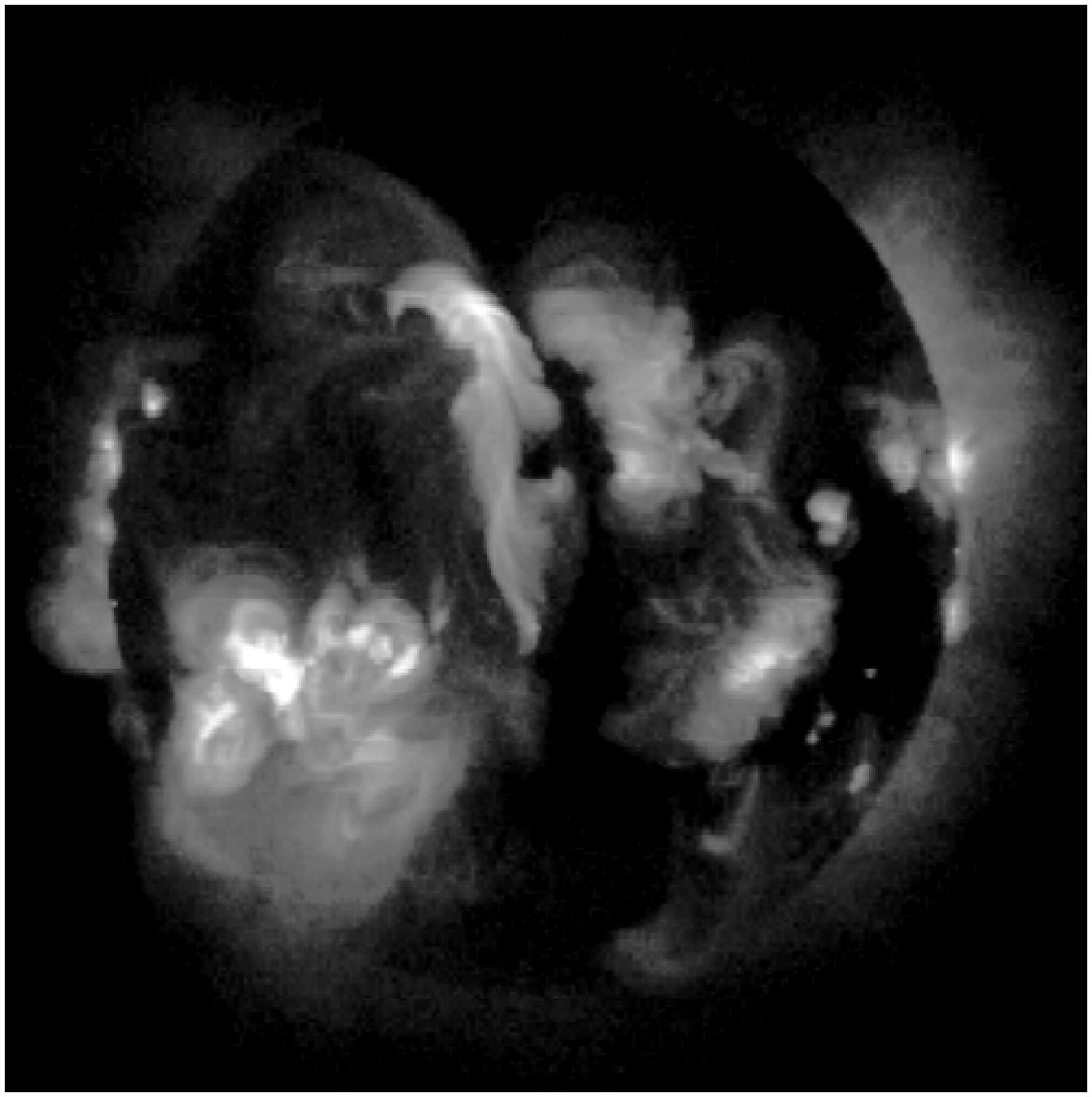}{0cm}{0}{26.2}{25.2}{-534}{-526} } 
\vskip 17.2truecm
\caption{Stellar X-ray eclipses and map reconstructions:  (a) YY Gem eclipse (G01a); 
(b) $\alpha$ CrB eclipse (G\"udel et al. 2003a).
Bottom, clockwise from upper left:   
AR Lac corona (Siarkowski et al. 1996);  $\alpha$ CrB corona (G\"udel et al. 2003a);
solar image for comparison (Yohkoh mission of ISAS, Japan); 
YY Gem corona (phase $\approx$ 0.875, G01a). }
\vskip -0.3truecm
\end{figure}

The emission is inhomogeneous, concentrated in active regions
in particular on the leading hemispheres of the stars (W90, O93) or between
the stars (Culhane et al. 1990; Siarkowski 1992; Siarkowski et al. 1996; Pre\'s, Siarkowski, \& Sylwester 1995) and at 
low to mid latitudes (W83, W86; W90; O93; G\"udel et al. 2001a = G01a; Fig. 2e), suggesting 
that the ``active''  filling factors  are modest 
(up to $10-20$\%). The possibility that some of the material links the two stars 
(Fig. 2c) opens up speculations on   
magnetic reconnection between the two coronae (see, e.g., 
Uchida \& Sakurai 1983). Such has also been suggested from radio eclipse observations of the
Algol-type binary  V505 Sgr (Gunn et al. 1999) and the pre-cataclysmic binary V471 Tau 
(Lim, White, \& Cully 1996).

It is not a priori clear in which type of structure large flares occur. For M dwarfs,
hydrodynamic simulations of large flares suggest active region sizes that cover a 
substantial fraction of the stellar surface (Reale et al. 2003).  VLBI results suggest
that flaring evolves from compact to extended structures, and this seems to be supported by
eclipse observations of flares on VW Cep and Algol  that indicate 
relatively compact emission sites with sizes of $\approx 0.85R_*$ (near one pole, Choi \& Dotani 1998),
$\approx 0.6R_*$ (scale height, near one pole, Schmitt \& Favata 1999) 
and 0.1$R_*$ (Schmitt, Ness, \& Franco 2003).  Densities of $5\times 10^{10}$ - several times
$10^{11}$~cm$^{-3}$ are found, similar to solar flares.

While the above examples are all extremely active stars,  
a transition to more solar-like behavior is expected as the activity declines.  A particularly favorable
configuration is offered by $\alpha$ CrB B, an intermediately old ($\approx 500$~Myr) 
solar analog that is totally eclipsed by an X-ray dark A star. Image reconstructions from light 
curves indicate a patchy structure with compact active regions, revealing
electron densities up to a few times $10^{10}$~cm$^{-3}$ but no evidence for an extended
corona (Schmitt \& K\"urster 1993; G\"udel et al. 2003a; Fig. 2).

\subsection{Rotational Modulation}

In the absence of eclipses, a more general approach to coronal structure is provided by monitoring
of rotational modulation. The Sun often shows rather pronounced X-ray rotational 
modulation as a small number of active regions alternately rotate into and out of view. However,  
rotational modulation is exceptional  among stars, one of the main 
reasons being that most of the brightest examples are highly active; such stars are
probably  covered with many active regions, and intense flaring often 
covers up low-amplitude modulations.  Such problems have hampered identification
of periodic X-ray modulation in the prototypical young active star, AB Dor
(White et al. 1996 and references therein; G\"udel et al. 2001b; but
see K\"urster et al. 1997), although radio observations reveal two emission 
peaks per rotation seen repeatedly over long time intervals.
They probably relate to preferred active longitudes  (Lim et al. 1992). 

Somewhat less active stars may be better-suited targets. EK Dra, a young,  
active solar analog, shows   rotational modulation both in X-rays and radio 
(G\"udel et al.  1995). The depth and length of the modulation constrains 
the X-ray coronal  height and the electron densities  ($> 3\times 10^{10}$~cm$^{-3}$) and 
demonstrates that much of the emitting material is concentrated in extended ``active regions''. 

The X-ray luminosity is also  below its possible maximum for stars  
in the ``supersaturation regime'' where the
X-ray luminosity decreases with increasing rotation velocity. The cause is not clear, but 
a deep modulation of the supersaturated young star VXR45 suggests 
that extreme activity in these stars is {\it not} due to complete coverage of the surface with 
active regions (Marino  et al. 2003).

Among  evolved stars,  the RS CVn binary HR~1099 (K1~IV + G5~V) 
has consistently shown X-ray/EUV rotational modulation, with a maximum at 
phases when the larger K subgiant is in front (Agrawal \& Vaidya 1988; 
Drake et al. 1994; Audard, G\"udel, \& Mewe 2001). The  rotationally modulated material 
can therefore be located - in contrast to results from 
eclipsing studies described above - on the K star hemisphere 
that {\it faces away} from the companion (Audard et al. 2001, using information
from Ayres et al. 2001, see \S 2.4).

\subsection{Doppler Information}

Doppler information from X-ray spectral lines may open up new ways of imaging 
stellar coronae as they rotate, or as they orbit around the center of gravity in binaries.
First attempts are encouraging although  the instrumental limitations are severe.
Ayres et al. (2001) find Doppler shifts in lines of HR~1099 that clearly agree
with the line-of-sight velocity of the subgiant K star. Periodic line broadening 
in YY Gem, on the other hand, suggests that both components are similarly X-ray luminous 
(G01a). Huenemoerder et al. (2003) find Doppler motions in
AR Lac to be compatible with coronae on both companions, close to the photospheric level.
For the  contact binary 44i Boo, Brickhouse et al. (2001) find periodic line 
shifts, and from the amplitudes and rotational modulation (Brickhouse \& Dupree 1998), 
they conclude that two dominant X-ray sources are
present, one very compact and one extended, both  located close to the stellar pole on the 
larger companion (see also Brickhouse, these proceedings).

\section{Energy Release and Coronal Structure}

The study of coronal structure confronts us with several problems that are difficult
to explain by scaling (up or down) of solar coronal structure: i) Characteristic 
coronal temperatures increase with increasing magnetic activity (Schrijver, Mewe, \& Walter 1984; 
Schmitt et al. 1990; G\"udel et al. 1997). ii) Characteristic coronal densities  increase analogously, 
at least in main-sequence stars (G\"udel et al. 2001b; Ness et al. 
2001, 2002). iii) The maximum X-ray luminosity exceeds the level expected from complete 
coverage of the surface with solar-like active regions by up to an order
of magnitude (Vaiana \& Rosner 1978). iv) Radio observations reveal  persistent 
high-energy non-thermal electrons in magnetically active stars even if
the lifetime of such a population should only be tens of minutes 
to about an hour  under coronal conditions (G\"udel 2002). Several of these features 
are reminiscent of flaring, as are some structural elements in stellar coronae (\S 2).
The remainder of this paper is devoted to the hypothesis that
coronal structure is intimately related to explosive energy release, i.e., flares. 

\subsection{Correlations between Coronal Emissions}

The time-averaged power from optical U-band flares correlates linearly with the low-level, 
``quiescent'' X-ray luminosity (Doyle \& Butler 1985;  Skumanich 1985).
The latter itself correlates with the rate of X-ray flares exceeding a given energy threshold (Audard
et al. 2000). Such would be expected if quiescent emission were physically related to the overall 
flaring rate in the active corona - for example if the detected X-ray flares represented the 
``tip of the iceberg'', the iceberg  being a distribution of flares of various amplitudes, 
the ocean being the combined emission of a large number of superimposed small flares. 

G\"udel \& Benz (1993) discuss a global relation between non-thermal
radio luminosities of active stars and their ``quiescent'' X-ray luminosities. Since the
lifetime of MeV electrons in coronal magnetic fields requires frequent acceleration,
a possible explanation again involves stochastic flares: the flare-accelerated
electrons could themselves act as the heating agents via chromospheric evaporation.
The smoking gun comes with the observation that the total radio and X-ray outputs
of solar flares follow the same correlation (Benz \& G\"udel 1994).

Since i) flares heat plasma, ii) increase the electron 
densities, therefore iii) increase the luminosity per surface area, and iv) amply produce 
non-thermal electrons, could it be that the observed phenomenology and structure on 
magnetically active stars is  due to an ensemble of stochastic flares?

\subsection{Coronal Variability: Quiescent versus Flares?}

Suggestive evidence has been gathered from light curves. A strong correlation 
between H$\gamma$ flare flux and simultaneous low-level X-ray flux in dMe stars suggests
that a large number of flare-like events may be responsible for the low-level emission (Butler et al.
1986). 
At least in main-sequence stars, there is little evidence for constant  emission
in high-sensitivity X-ray data (Audard et al. 2003). Frequent faint, flare-like X-ray fluctuations (Fig. 3b) are often 
accompanied - in fact slightly preceded - by optical (U band) bursts, the latter being a signature
of the initial bombardment of the chromosphere by high-energy electrons (G\"udel et al. 2002).
Stochastic chromospheric evaporation could thus be the process that enriches the corona 
with plasma. In any case, 
{\it any successful model of coronal structure and heating must account for the
observed continuous variability.} 
 
\subsection{Coronal Flare Statistics}

The suggestion that stochastically occurring flares may be largely responsible for
coronal heating is known as the ``microflare'' or ``nanoflare'' hypothesis in solar 
physics (Parker 1988). Observationally, it is supported by evidence  for the presence of
numerous  small-scale flare events occurring in the solar corona 
at any time (e.g., Lin et al. 1984). Their distribution in energy is  a 
power law,
\begin{equation}\label{e:powerlaw}
\frac{dN}{dE} = k E^{-\alpha} 
\end{equation}
where $dN$ is the number of flares (per unit time) with a total 
energy in the  interval [$E,E+dE$]. If $\alpha\ge 2$, then the energy integration (for a given 
time interval) diverges for $E_{\rm min} \rightarrow 0$, i.e., by 
extrapolating the power law to sufficiently small flare energies, {\it any} 
energy release power can be attained. This is not the case for $\alpha <2$.
Evidently, then, one needs to measure the energy distribution of 
a statistically relevant number of flares. Solar studies have repeatedly
resulted in $\alpha$ values of the order of 1.6--1.8 for ordinary
solar flares (Crosby, Aschwanden, \& Dennis 1993), but recent studies of low-level  flaring
suggests $\alpha = 2.0 - 2.6$ (Krucker \& Benz 1998; Parnell \& Jupp 2000). 

Relevant stellar studies have been rare. Early results converged to $\alpha < 2$  for
stellar samples (Table 1).
Mixing  stars at different distances and with different luminosities 
can introduce  strong statistical bias. To avoid this,
Audard, G\"udel, \& Guinan (1999) and Audard et al. (2000)   applied a flare search 
algorithm to {\it EUVE} light curves of individual active main-sequence stars, taking 
into account flare superpositions and
various binning to recognize weak flares. These results indicate a predominance of 
relatively steep power laws including $\alpha \ge 2$.

The identification of weak flares close to the low-level emission 
level is an ill-defined problem. Limited signal-to-noise ratios add to the problem.  
Some of these complications were overcome by fully forward modeling 
the superposition of a statistical ensemble of X-ray or EUV flares (Kashyap et al. 2002; 
G\"udel et al. 2003b) including an analytical formulation (Arzner \& G\"udel 2003). 
The results of these investigations are in full agreement, converging to $\alpha \approx 2.0 - 2.5$
for M dwarfs. If the power-law flare energy distribution extends by about 1--2 orders of magnitude below
the actual detection limit in the light curves, then the {\it entire} emission could be explained
by stochastic flares. The coronal heating process would thus be one solely due to {\it time-dependent}
heating by flares.

\begin{table}[t!] 
\caption{Stellar radiative flare energy distributions} 
\begin{tabular}{llll} 
\tableline 
Star sample	       & Photon energies      & $\alpha$       & References \\
\tableline 
M dwarfs	       &  0.05--2~keV	      & 1.52$\pm 0.08$ & Collura et al. (1988) \\    
M dwarfs	       &  0.05--2~keV	      & 1.7$\pm 0.1$   & Pallavicini et al. (1990) \\
RS CVn binaries        & EUV		      & 1.6	       & Osten \& Brown (1999)\\
Two G dwarfs	       & EUV		      & 2.0--2.2       & Audard et al. (1999)\\
F-M dwarfs	       & EUV		      & 1.8--2.3       & Audard et al. (2000)\\
Three M dwarfs         & EUV		      & 2.2--2.7       & Kashyap et al. (2002)\\
AD Leo  	       & EUV\&0.1--10~keV     & 2.0--2.5       & G\"udel et al. (2003b) \\
AD Leo  	       & EUV                  & $2.3\pm 0.1$   & Arzner \& G\"udel (2003) \\
\tableline\tableline
\end{tabular}
\end{table}

\subsection{Observables of Stochastic Flaring}

In order for stochastic flaring to be an acceptable coronal heating  mechanism, a number of 
observables should be correctly reproduced. Flares develop characteristically in 
emission measure (EM) and temperature $T$ (fast rise to peak, slow decay). The individual flare 
histories cannot be resolved in the data but the statistical, time-integrated thermal 
histories of the ensemble of flares reflect in the time-averaged
differential EM (DEM) distribution. The 
flares follow a relation in which the peak luminosity (or EM,  hence  the emitted X-ray
energy) is a function of  the peak $T$ (Feldman, Laming, \& Doschek 1995).
For a simple flare shape,  integration of this dependence over
time and over the flare energy distribution yields an analytic expression for the DEM:
Below the turnover,  DEM $= n^2 dV/ d\mathrm{log}T \propto T^{2\tau_T/\tau_n}$ where
$\tau_T$ and $\tau_n$ are the (e-folding) decay times of the flare $T$ and
electron density, respectively ($\tau_T/\tau_n = 0.5 - 2$, Reale et al. 1993). Above the turnover, 
apart from small correction terms, DEM $\propto   T^{-b(2-\alpha)}$ where $b \approx 5 \pm 1$ 
is derived from the Feldman et al. relation. Such DEMs
are reminiscent of those found from X-ray spectroscopy (Telleschi et al., these proc).

\begin{figure}[t!]
\vskip 5truecm
\hbox{\plotfiddle{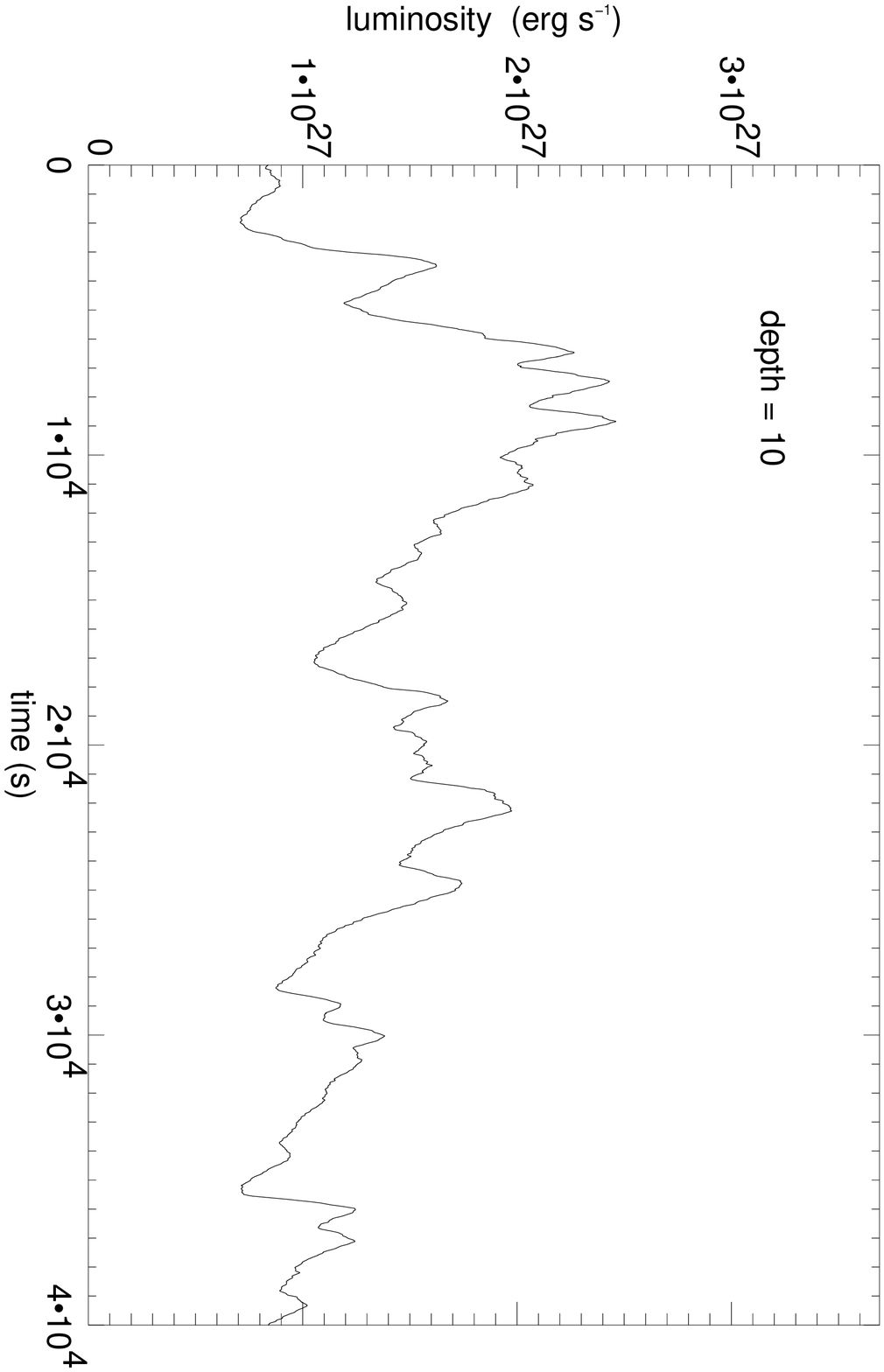}{0cm}{90}{25}{25}{-195}{10} \plotfiddle{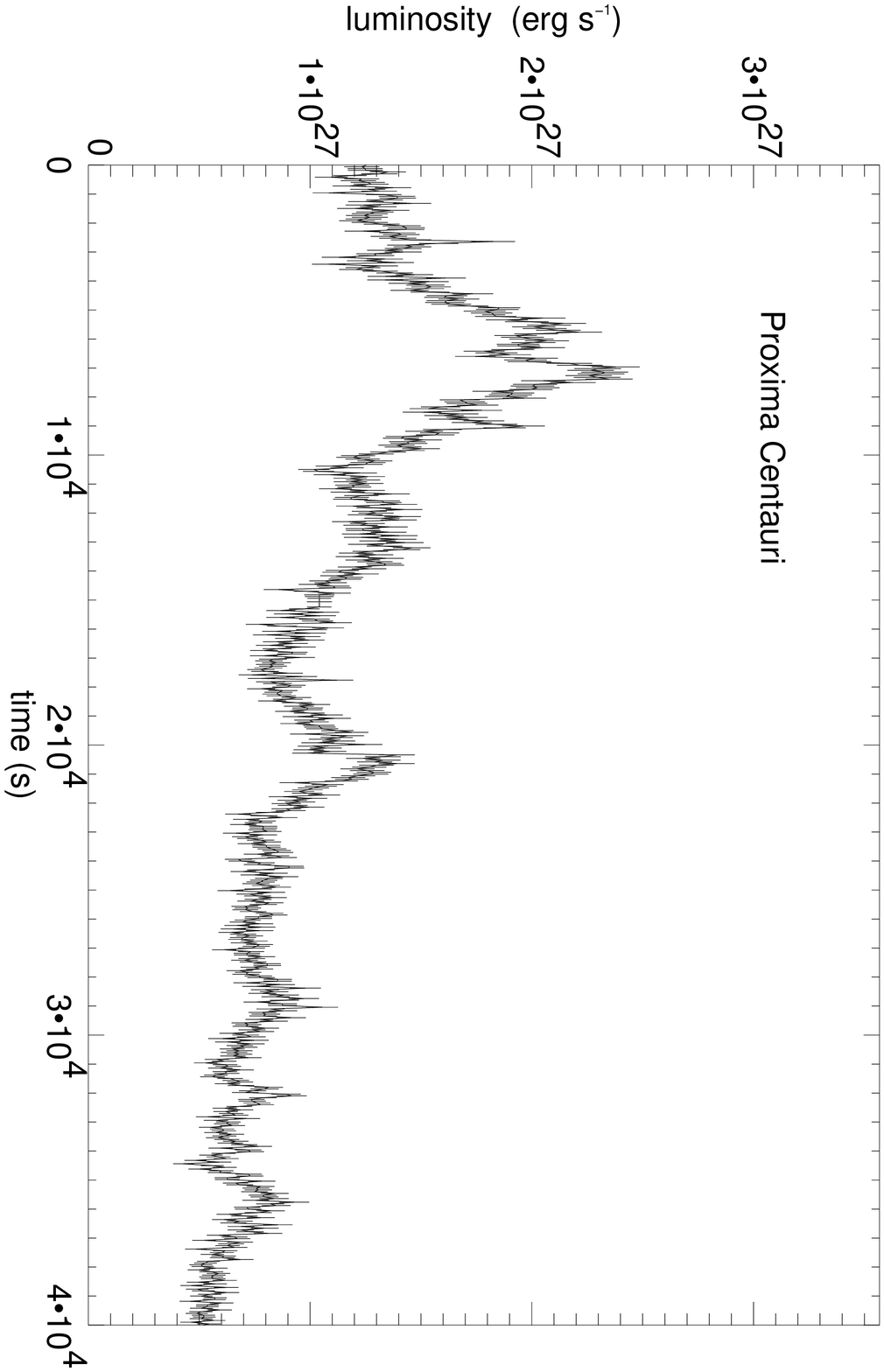}{0cm}{90}{25}{25}{-390}{10}}
\vskip 4truecm
\hbox{\plotfiddle{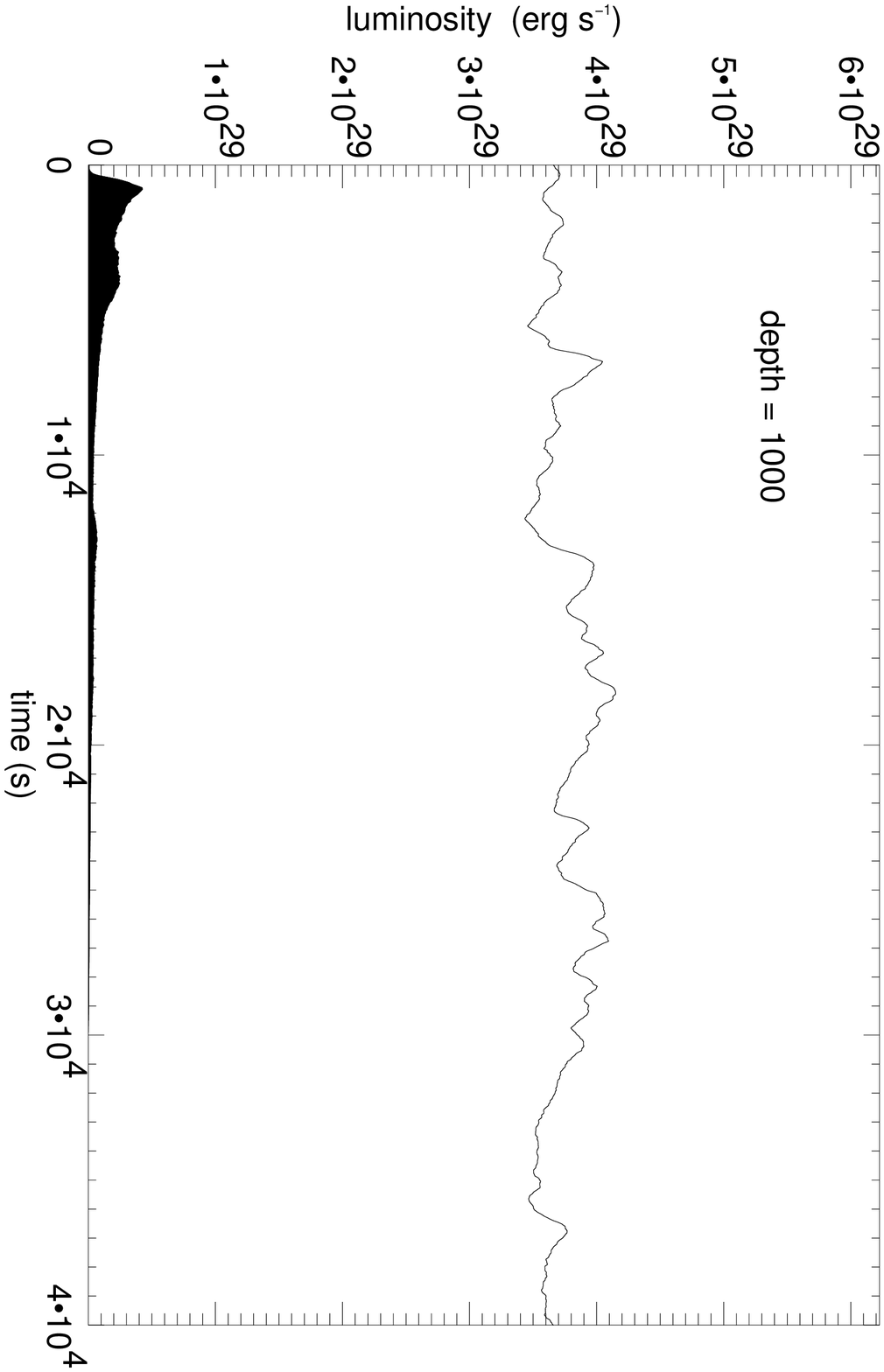}{0cm}{90}{25}{25}{-195}{10} \plotfiddle{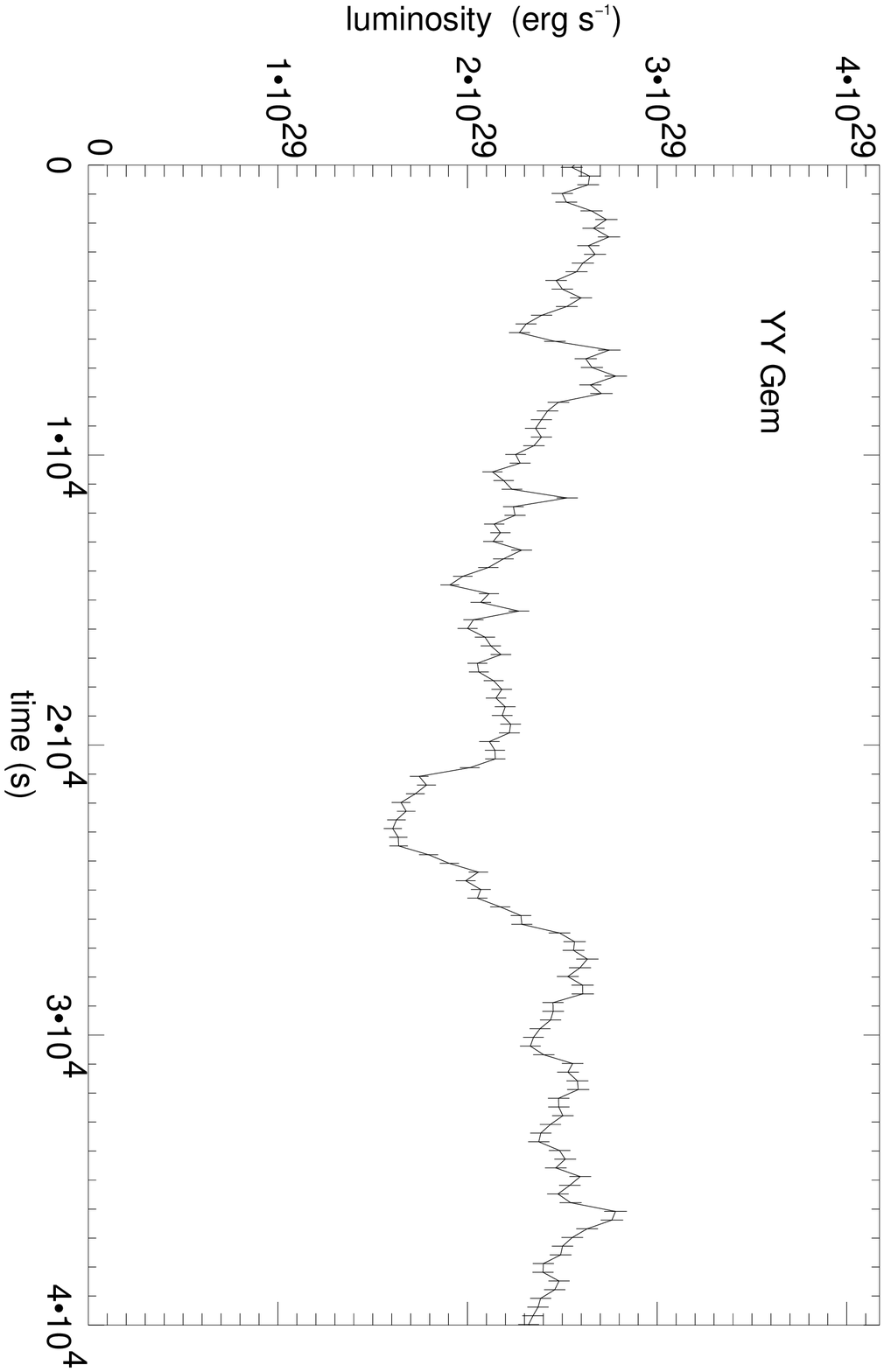}{0cm}{90}{25}{25}{-390}{10}}
\vskip -0.8truecm
\caption{X-ray light curves. {\it Left panels:} Synthetic light curves from superimposed
flares; largest to smallest amplitude = 10 (upper) and 1000 (lower plot). A
flare on Proxima Centauri (G\"udel et al. 2002) has been used as a template, 
shown filled in the lower plot for the  maximum amplitude
contributing to the light curve. {\it Right panels:} Observed X-ray light curves  for comparison. Proxima
Centauri (upper, G\"udel et al. 2003c) and YY Gem (lower plot, including eclipse; after G01a).}
\end{figure}

Flares are often defined in light curves as excess emission above an envelope
that itself is defined as ``quiescent emission''. This definition is problematic
for two reasons. i) It is strongly dependent on the achieved signal-to-noise ratio. Any light
curve shows quiescent characteristics at sufficiently small S/N or time resolution. 
ii) The superposition of stochastically occurring flares equally produces a lower-envelope 
emission level that is always present. 

Fig. 3 shows simulations of superimposed flares  
with $\alpha = 2.2$, compared with observations. The large flare
observed on Proxima Centauri (G\"udel et al. 2003c) was used as a shape template.
The first example uses flare energies spread over a factor of ten only, the second uses three
orders of magnitude. In the latter case, the curve has smoothed out to an extent that it is
dominated by ``quiescent'' emission. The individual peaks are merely the peaks of the most energetic
flares  in the ensemble. The available stellar area may constrain the energy range of stochastic
flares; in small, late-type M dwarfs, flaring active regions may cover a significant
fraction of the surface (Reale et al. 2003), limiting the number
of simultaneous flares (upper example in Fig. 3, e.g., Proxima Centauri).

Spectroscopic density measurements of a time-integrated, stochastically flaring corona 
should then yield values equivalent to the values derived from time-integrating the
spectrum over a large flare. This is indeed the case. While large stellar X-ray flares
achieve peak electron densities of several times $10^{11}$~cm$^{-3}$,
the time-integrated X-ray spectrum of the Proxima Centauri flare described in 
G\"udel et al. (2002, 2003c) yields a characteristic density of log$n_e \approx 10.5 \pm 0.25$
derived from O~VII, which compares favorably with densities found for magnetically
active main-sequence stars (Ness et al. 2001, 2002).

\section{Conclusions}

Magnetically active stars frequently show 
{\it bi-modal coronal structure:} i) Compact ``active
regions'' at small heights appear to be ``solar-like'' but often require pressures and densities comparable
to solar flare pressures/densities; some compact sources have explicitly been seen to develop into
expanding flares in radio VLBI observations.  ii) Very extended, globally well ordered 
magnetic structure seen in X-ray eclipses (as  hot gas) or in radio VLBI (as a non-thermal electron
population) reaches size scales of several  times the stellar radius. These structures
may be related to cooling, expanding  ``flare remnants'' or large, van Allen-Belt like closed
magnetospheric structures. 
Their thermal, density, and temporal properties  are unlike typical structures in 
the non-flaring Sun but are reminiscent of flaring plasma. We have discussed expected properties of
a stochastically flaring corona, such as light curves, the DEM, and the 
 characteristic densities. While these predicted observables deviate considerably from
values of the non-flaring Sun, they are reminiscent of observations of magnetically active stars.

Why should enhanced flaring occur on magnetically active stars, or in other words, 
why should stars with an enhanced level of magnetic field production be prone to develop
flare-like magnetic coronae?  We speculate that, as the magnetic dynamo 
increases toward more active stars,  the surface filling factor of strong magnetic fields 
increases, increasing also the volume filling factor of closed loops in the corona. As the 
magnetic loops are packed closer together than on the Sun, they interact more frequently, producing
a higher rate of field annihilation and reconnection, thus generating flares
at a higher rate. These energy release events liberate magnetic energy and 
heat chromospheric plasma, evaporating it by overpressure into the
magnetic loops. One thus expects both a larger average density in the
corona and a higher temperature if the flare rate is higher. Increased densities and temperatures 
both are, in this picture, a natural consequence of a denser packing of magnetic fields
in the corona, induced by a more efficient dynamo in stars with more rapid 
rotation. The determining factor of the coronal temperature and emission measure (hence the luminosity)
is thus the flare frequency (above a certain energy threshold), which in turn is a consequence of the
amount of magnetic flux production in the stellar interior.

\acknowledgments I thank Philippe Andr\'e, Arnold Benz, Robert Mutel, and Marek Siarkowski for
                 providing figures reproduced in this paper.

\end{document}